\documentclass[12pt]{article}
\usepackage{amsfonts}
\newcommand{\D}{\displaystyle}
\begin{document}

\author{{\bf A. B\'erard$^{a}$, J. Lages$^{b}$ and H. Mohrbach$^{c}$ }\\
\\
\textit{$^{a}$ L.P.M.C.T., Institut\ de\ Physique, } \\
\textit{1 blvd. Fran\c{c}ois Arago, 57070, Metz, France} \\
\\
\textit{$^b$ Groupe de Physique Th\'eorique,} \\
\textit{Laboratoire de Physique Quantique,} \\
\textit{UMR 5626 du CNRS, Toulouse, France} \\
\\
\textit{$^c$ Institut Charles Sadron,} \\
\textit{CNRS UPR 022,} \\
\textit{6 rue Boussingault, 67083 Strasbourg Cedex, France}}
\title{\bf Restoration of Angular Lie Algebra Symmetries from a 
Covariant Hamiltonian}
\date{1 october 2001}
\maketitle

\begin{abstract}
The sO(3) and the Lorentz algebra symmetries breaking with gau\-ge curvatures a
re studied by means of a covariant Hamiltonian. The restoration of these
algebra symmetries in flat and curved spaces is performed and led to the 
apparition of a monopole field. Then in the context of the  Lorentz algebra
we consider an application to the gravitoelectromagnetism theory.
In this last case a
qualitative relation giving a mass spectrum for dyons is established.
\end{abstract}

\section{Introduction}

The concept of symmetry breaking is fundamental in science and particularly
in physics. The breaking of the Lorentz algebra symmetry by gauge curvature
has been studied in a recent paper \cite{NOUS2} from a covariant
Hamiltonian, defined in a tangent bundle frame. The restoration of this
algebra symmetry is possible by introducing a Poincar\'{e} momentum 
\cite{POINCARE} and as a consequence a Dirac monopole \cite{DIRAC}.

In the present paper we insist further on the link between the restoration
of algebra symmetry and the appearance of monopoles, similarly to the
case of the duality symmetry of electromagnetism 
which is the symmetry of the free Maxwell equations under rotations 
of electric and magnetic fields in U(1) gauge theories.
Here we also extend this approach \cite{NOUS2} to the case of the general
relativity. The section 2 recalls the basic formalism of our
approach \cite{NOUS2,NOUS1} and deepens it by means of a generalization of
Poisson brackets taking gauge fields into account. In the section 3 we
study the restoration of the sO(3) algebra symmetry in a flat and a curved
space, with and without the introduction of the connexion field. In the last
section we consider the restoration of the Lorentz algebra symmetry in a 
curved space and its possible application to the theory of 
gravitoelectromagnetism where a qualitative relation giving a mass spectrum 
for dyon is deduced.

We want to remark that the brackets we use in our formalism is connected 
to those introduced by Feynman in his remarkable demonstration of the 
Maxwell equations where he tried to develop a quantization procedure 
without resort to a Lagrangian or a
Hamiltonian. These ideas are exposed by Dyson 
in an elegant publication \cite{DYSON}. 
The interpretation of the Feynman's derivation
of the Maxwell equations has aroused a great interest among the
physicists. In particular Tanimura \cite{TANIMURA} has generalized Feynman's
derivation to a Lorentz covariant form with a scalar evolution 
parameter. An extension of the Tanimura's approach has been provided 
\cite{NOUS3}  in using the Hodge duality in 
order to derive the
two groups of Maxwell equations with a magnetic monopole in flat and
curved spaces. For his part Lee \cite{LEE} has included the
descriptions of relativistic and non relativistic particles in an
electromagnetic field, not long after Chou \cite{CHOU} has established a
dynamical equation for spinning particles. A rigorous mathematical
interpretation of Feynman's derivation connected to the inverse problem for the
Poisson dynamic has been made by Carinena {\it et al} \cite{CARINENA}. Then
Hojman and Shepley  \cite{HOJMAN} and Hughes \cite{HUGHES} have placed this
approach in the context of the Helmholtz's inverse problem of the calculus
of variations. Finally we have also to note the recent works of Montesinos and
P\'erez-Lorenzana \cite{MONTESINOS}, Singh and Dadhich \cite{SINGH} and
Silagadze \cite{SILAGADZE} which provide new looks on the Feynman's derivation.

\section{Basic formalism}

Let $M$ be a three dimensional vectorial manifold where the set of vector
components is homeomorphic to $\mathbb{R}^{3}$. In the following of this paper
we shall work with these $\mathbb{R}$-triplets (or $\mathbb{R}$-quadruplets 
in the
case of gravitoelectromagnetism in the last part).
We introduce a non commutative algebra by means of skew symmetric brackets
with the following distributive properties 
\begin{equation}
[ x^{i},y^{j}z^{k}]=[x^{i},y^{j}]z^{k}+[x^{i},z^{k}]y^{j}
\mbox{ , and
 }[x^{i}y^{j},z^{k}]=[y^{j},z^{k}]x^{i}+[x^{i},z^{k}]y^{j},
\end{equation}
and we require the local property 
\begin{equation}\label{local}
\left[ x^{i},x^{j}\right] =0.
\end{equation}
Let $\tau $ be the parameter of the group of diffeomorphisms $g$ 
\begin{equation}
g(\mathbb{R}\times M_{3})\longrightarrow M_{3}:g(\tau ,x^{i})=g^{\tau
}x^{i}=x^{i}(\tau ),
\end{equation}
then the ''velocity vector'', $\dot{x}^{i}=\frac{d}{d\tau }g^{\tau
}x^{i}$, associated to the particle of mass $m$ is naturally introduced by
the dynamic equation $\dot{x}^{i}=\frac{dx^{i}}{d\tau }=\left[
x^{i},H\right] $, where the Hamiltonian $H$ is a priori an expandable
function of $x^{i}$ and $\dot{x}^{i}.$
Under these conditions we obtain the two relations 
\begin{equation}
\left\{ 
\begin{array}{c}
H=\frac{1}{2}mg^{ij}(x,\dot{x})\dot{x}_{j}\dot{x}
_{i}, \\ 
\\ 
m\left[ x^{i},\dot{x}^{j}\right] =g^{ij}(x,\dot{x}),
\end{array}
\right.   \label{metrique}
\end{equation}
where $g^{ij}(x,\dot{x})$ is the metric tensor of the ''physical
space'' which is chosen for the moment as an Euclidean flat space $g^{ij}(x,
\dot{x})=\delta ^{ij}$. Our construction shows clearly that this
Hamiltonian is covariant in a similar way to the four dimensionnal covariant 
Hamiltonian introduced by Goldstein \cite{GOLDSTEIN} 
in the context of electromagnetism.

From our definitions we derive the following relation for expandable
functions 
\begin{equation}
\left[ f(x,\dot{x}),g(x,\dot{x})\right] =\left\{
f(x,\dot{x}),g(x,\dot{x})\right\} +\left[ \dot{x}
^{i},\dot{x}^{j}\right] \frac{\partial f(x,\dot{x})}{
\partial \dot{x}^{i}}\frac{\partial h(x,\dot{x})}{\partial 
\dot{x}^{j}},
\end{equation}
where we have used the usual Poisson brackets formalism but acting here on
functions defined on the tangent bundle space 
\begin{equation}
\left\{ f(x,\dot{x}),g(x,\dot{x})\right\} =
\frac{1}{m}\left(
g^{ij}\frac{
\partial f(x,\dot{x})}{\partial x^{i}}\frac{\partial h(x,
\dot{x})}{\partial \dot{x}^{j}}-g^{ij}\frac{\partial f(x,\dot{x}
)}{\partial \dot{x}^{i}}\frac{\partial h(x,\dot{x})}{
\partial x^{j}}\right).
\end{equation}
We also introduce the following notation
\begin{equation}
J(f,g,h)=\left[ f,\left[ g,h\right] \right] +\left[ g,\left[ h,f\right]
\right] +\left[ h,\left[ f,g\right] \right]
\end{equation}
in order to help us to describe the structure of the Jacobi identities.

It is easy to check that, for a particle with an electric charge $q$, the
tensor $\left[ \dot{x}^{i},\dot{x}^{j}\right] $ 
is a skew symmetric tensor and will be noted $\frac{q}{m^2}{\mathcal F}
^{ij}(x,\dot{x})$. It corresponds to the generalization of the only position
dependent three dimensional electromagnetic tensor introduced 
in a preceding paper \cite{NOUS1}.
Thus in the presence of such a gauge curvature 
${\mathcal F}^{ij}(x,\dot{x})$ we obtain for the Jacobi structures
\begin{equation}
\left\{ 
\begin{array}{l}
J\left( f(x),g(x),h(x)\right)=0,\\
J\left( f(x),g(x),h(x,\dot{x})\right)=0, \\ 
J\left( f(x),g(x,\dot{x}),h(x,\dot{x})\right)=\D\frac{q}{m^{3}
}\frac{\partial f}{\partial x^{i}}\frac{\partial g}{\partial \dot{x}
^{j}}\frac{\partial h}{\partial \dot{x}^{k}}\frac{\partial {\mathcal F
}^{jk}}{\partial \dot{x}_{i}}, \\ 
J\left( f(x,\dot{x}),g(x,\dot{x}),h(x,\dot{x})\right)=
\D\frac{q}{m^{3}}\left( \frac{\partial {\mathcal F}^{ij}}{\partial x_{k}}+
\frac{\partial {\mathcal F}^{jk}}{\partial x_{i}}+\frac{\partial {\mathcal F}
^{ki}}{\partial x_{j}}\right) \frac{\partial f}{\partial \dot{x}^{i}}
\frac{\partial g}{\partial \dot{x}^{j}}\frac{\partial h}{\partial 
\dot{x}^{k}}\\ 
+\D\frac{q}{m^{3}}\left( \frac{\partial f}{\partial x^{i}}\frac{\partial g}{
\partial \dot{x}^{j}}\frac{\partial h}{\partial \dot{x}^{k}}+
\frac{\partial f}{\partial \dot{x}^{k}}\frac{\partial g}{\partial
x^{i}}\frac{\partial h}{\partial \dot{x}^{j}}+\frac{\partial f}{
\partial \dot{x}^{j}}\frac{\partial g}{\partial \dot{x}^{k}}
\frac{\partial h}{\partial x^{i}}\right) \frac{\partial {\mathcal F}^{jk}}{
\partial \dot{x}_{i}} \\ 
+\D\frac{q^2}{m^4}
\left( {\mathcal F}^{kl}\frac{\partial {\mathcal F}^{ij}}{\partial 
\dot{x}^{l}}+{\mathcal F}^{il}\frac{\partial {\mathcal F}^{jk}}{\partial 
\dot{x}^{l}}+{\mathcal F}^{jl}\frac{\partial {\mathcal F}^{ki}}{
\partial \dot{x}^{l}}\right) \frac{\partial f}{\partial 
\dot{x}^{i}}\frac{\partial g}{\partial \dot{x}^{j}}\frac{\partial h}{
\partial \dot{x}^{k}}.
\end{array}
\right.   \label{jacobi}
\end{equation}
The first and the second one of these relations (\ref{jacobi}) 
are the usual Jacobi identities and express here the local property
(\ref{local}) in a Euclidean flat space $g^{ij}=\delta^{ij}$. 
The others express the existence of the gauge curvature and 
are trivially null in its absence.

It is also possible to show that the Hamiltonian is invariant under
his dynamical equation.The 
equation of motion of a particle is then simply obtained by writing
\begin{equation}
m\ddot{x}^{i}=m\left[ \dot{x}^{i},H\right] =q{\mathcal F}
^{ij}(x,\dot{x})\dot{x}_{j}.
\end{equation}

In order to study the symmetry breaking of the sO(3) algebra
we introduce the usual angular momentum
$L^{i}=m\varepsilon ^{i}{}_{jk}x^{j}\dot{x}^{k}$
which is a constant of the motion in absence of gauge field.

\section{sO(3) algebra}

One of the most important symmetry which occur in physical problems is
naturally the spherical symmetry corresponding to the isotropy of the
physical space which is connected to the sO(3) algebra. In the following we
show that this symmetry is broken when an electromagnetic field is introduced.

\subsection{sO(3) algebra without gauge curvature}

No electromagnetic field implies $\left[ \dot{x}^{i},\dot{x}
^{j}\right] =0$, and the sO(3) Lie algebra defined with our brackets is then
equal to the standard algebra defined in terms of our Poisson brackets
\begin{equation}
\left\{ 
\begin{array}{l}
[ x^{i},L^{j}]=\left\{ x^{i},L^{j}\right\}=\varepsilon ^{ijk}x_{k},
\\ \\
\left[\dot{x}^{i},L^{j}\right]=\left\{ \dot{x}^{i},L^{j}\right\}
=\varepsilon ^{ijk}\dot{x}_{k}, \\ \\
\left[ L^{i},L^{j}\right]=\left\{ L^{i},L^{j}\right\}=\varepsilon ^{ijk}L_{k}.
\end{array}
\right. 
\end{equation}

\subsection{sO(3) algebra with gauge curvature}

By choosing $\left[ \dot{x}^{i},\dot{x}^{j}\right] =\frac{q
}{m^{2}}{\mathcal F}^{ij}(x,\dot{x})$, where the field ${\mathcal F}
^{ij}$ is \textit{a priori} position and velocity dependent,
we generalize the gauge theories studied recently with only position 
dependent fields \cite{NOUS2,NOUS1}.

So trying to retrieve the sO(3) Lie algebra we find 
\begin{equation}
\left\{ 
\begin{array}{l}
\D\left[ x^{i},L^{j}\right]=
\left\{ x^{i},L^{j}\right\} =\varepsilon ^{ijk}x_{k},
\\ 
\\ 
\D\left[ \dot{x}^{i},L^{j}\right]=\left\{ \dot{x}^{i},L^{j}\right\}
+\frac{q}{m}\varepsilon ^{j}{}_{kl}x^{k}{\mathcal F}^{il}(x,\dot{x})
\\ 
\D\phantom{\left[ \dot{x}^{i},L^{j}\right]}=\varepsilon ^{ijk}\dot{x}_{k}+\frac{q}{m}\varepsilon
^{j}{}_{kl}x^{k}{\mathcal F}^{il}(x,\dot{x}), \\ 
\\ 
\D\left[ L^{i},L^{j}\right]=\left\{ L^{i},L^{j}\right\} +q\varepsilon
^{i}{}_{kl}\varepsilon ^{j}{}_{ms}x^{k}x^{m}{\mathcal F}^{ls}(x,
\dot{x}) \\ 
\D\phantom{\left[ L^{i},L^{j}\right]}
=\varepsilon ^{ijk}L_{k}+q\varepsilon ^{i}{}_{kl}\varepsilon
^{j}{}_{ms}x^{k}x^{m}{\mathcal F}^{ls}(x,\dot{x}).
\end{array}
\right. 
\end{equation}
The sO(3) algebra is then broken by the gauge curvature and we propose to
recover the standard algebra in two manners.

\subsubsection{Without connection}

First we use a ''transformation law'' which generalizes angular momentum 
\begin{equation}
L^{i}\rightarrow {\mathcal L}^{i}=L^{i}+{{\mathcal M}}^{i}(x,\dot{x}),
\end{equation}
and we naturally require the usual algebra for this new angular momentum 
\begin{equation}
\left\{ 
\begin{array}{l}
\left[ x^{i},{\mathcal L}^{j}\right]=\left\{ x^{i},{\mathcal L}^{j}\right\} =\varepsilon
^{ijk}x_{k}, \\ 
\\ 
\left[ \dot{x}^{i},{\mathcal L}^{j}\right]=\left\{ \dot{x}^{i},{\mathcal L}
^{j}\right\} =\varepsilon ^{ijk}\dot{x}_{k}, \\ 
\\ 
\left[ {\mathcal L}^{i},{\mathcal L}^{j}\right]=\left\{ {\mathcal L}^{i},{\mathcal L}^{j}\right\}
=\varepsilon ^{ijk}{\mathcal L}_{k}.
\end{array}
\right.   \label{structure}
\end{equation}
From the first relation in (\ref{structure}) we deduce that 
\begin{equation}
{{\mathcal M}}^{i}(x,\dot{x})=M^{i}(x),
\end{equation}
whereas the second 
\begin{equation}
\left[ \dot{x}^{i},M^{j}\right]=-\frac{1}{m}\frac{\partial M^{j}(x)}{
\partial x_{i}}=-\frac{q}{m}\varepsilon ^{j}{}_{kl}x^{k}{\mathcal F}^{il}(x,
\dot{x}),
\end{equation}
implies the important property that the gauge curvature is velocity
independent 
\begin{equation}
{\mathcal F}^{ij}(x,\dot{x})=F^{ij}(x).
\end{equation}
This result is a consequence of the sO(3) algebra in a flat space, it is
different when we consider a curved space as we shall discuss later.

The third one gives 
\begin{equation}
M^{i}=\frac{1}{2}q\varepsilon _{jkl}x^{i}x^{k}F^{jl}(x)=-q\left( 
\overrightarrow{r}.\overrightarrow{B}\right) x^{i},  \label{momentum}
\end{equation}
where we introduce the magnetic field $\overrightarrow{B}$ that
must have the same form as the Dirac magnetic monopole field 
\begin{equation}
\overrightarrow{B}=\frac{g}{4\pi }\frac{\overrightarrow{r}}{r^{3}},
\end{equation}
in order to be a solution of the second equation in (\ref{structure}).

The new quantity ${M}$ introduced in (\ref{momentum}) is the well known
Poincar\'{e} momentum \cite{POINCARE} already deduced in a preceding paper 
\cite{NOUS2} and the total angular momentum is then defined by 
\begin{equation}
\overrightarrow{{\mathcal L}}=\overrightarrow{L}-\left( \overrightarrow{r}.
\overrightarrow{B}\right) \overrightarrow{r}.
\end{equation}

We note that in the precedent derivation 
the following Jacobi identity is directly realized
\begin{equation}
J\left( x^{i},\dot{x}^{j},\dot{x}^{k}\right) =
\D\frac{q}{m^3}\frac{\partial
F^{jk}(x)}{\partial \dot{x}_{i}}=0,
\end{equation}
whereas if we impose the third Jacobi identity 
\begin{equation}
J\left( \dot{x}^{i},\dot{x}^{j},\dot{x}^{k}\right) =
\D\frac{q}{m^3}
\left(
\frac{\partial F^{ij}(x)}{\partial x_{k}}+\frac{\partial F^{jk}(x)}{\partial
x_{i}}+\frac{\partial F^{ki}(x)}{\partial x_{j}}
\right)=0,
\end{equation}
we retrieve the Bianchi equation that is the first group of Maxwell
equations for three dimensional electromagnetic field without monopole (the
Poincar\'{e} momentum is equal to zero, the monopoles are absorbed by
anti-monopoles).

\underline{Remark:} The Poincar\'{e} momentum can also be seen to be related
to the Wess-Zumino term introduced by Witten in the case of a simple
mechanical problem \cite{WITTEN}. Indeed let a physical system, with a
spatial-temporal reflection symmetry, formed by a particle of mass $m$
constrained to move on a circle. The result is that this particle is
submitted to a strength having the following form $qg\varepsilon _{ijk}x_{k}
\dot{x}_{j}$ , that can be understood as a Lorentz force acting on
an electric charge $q$, due to its interaction with a magnetic monopole of
magnetic charge $g$ located at the centre of the circle. For the quantum
version of this system, Witten has retrieved the Dirac quantization
condition by means of topological techniques.

\subsubsection{With connection}

In this section we consider two ''transformation laws'', one corresponding
to the angular momentum and the other corresponding to the velocity 
\begin{equation}
\left\{ 
\begin{array}{c}
L^{i}\rightarrow {{\mathcal L}}^{i}=L^{i}+{{\mathcal M}}^{i}(x,
\dot{x}), \\ 
\\ 
\dot{x}^{i}\rightarrow p^{i}=m\dot{x}^{i}+q{\mathcal A}^{i}(x,
\dot{x}).
\end{array}
\right. 
\end{equation}
The second transformation law is nothing else than a Legendre transformation
which is defined from the tangent bundle space to the cotangent bundle
space. The bracket between the position and the linear momentum plays now
the role of a ''second metric tensor'' $G$ of the space. 

We get instead of (\ref{metrique}) 
\begin{equation}
\left[ x^{i},p^{j}\right]=G^{ij}(x,\dot{x})=g^{ij}+\frac{q}{m}\frac{
\partial {\mathcal A}^{j}(x,\dot{x})}{\partial \dot{x}^{i}}.
\end{equation}
In this section we consider only the case where $G^{ij}=\delta ^{ij}$ ,
then ${\mathcal A}^{i}(x,\dot{x})=A^{i}(x)$, implying that the
brackets between two gauge fields is zero. The theory is abelian in
the sense that $\left[ A^{i}(x),A^{j}(x)\right] =0$.

In the goal to rescue the sO(3) symmetry we want the following relations to
be realized 
\begin{equation}
\left\{ 
\begin{array}{l}
\left[ p^{i},p^{j}\right]=\left\{ p^{i},p^{j}\right\} =0, \\ 
\\ 
\left[ x^{i},{{\mathcal L}}^{j}\right]=\left\{ x^{i},{{\mathcal L}
}\right\} =\varepsilon ^{ijk}x_{k}, \\ 
\\ 
\left[ p^{i},{{\mathcal L}}^{j}\right]=\left\{ p^{i},{{\mathcal L}
}^{j}\right\} =\varepsilon ^{ijk}p_{k}, \\ 
\\ 
\left[ {{\mathcal L}}^{i},{{\mathcal L}}^{j}\right]=\left\{ 
{{\mathcal L}}^{i},{{\mathcal L}}^{j}\right\} =\varepsilon
^{ijk}{{\mathcal L}}_{k}.
\end{array}
\right.   \label{structure2}
\end{equation}
In (\ref{structure2}), the first relation gives 
\begin{eqnarray}
{\mathcal F}^{ij}(x,\dot{x}) &=&m\left[ \dot{x}
^{j},A^{i}(x)\right] -m\left[ \dot{x}^{i},A^{j}(x)\right]   \nonumber
\\
&=&\frac{\partial A^{j}(x)}{\partial x^{i}}-\frac{\partial A^{i}(x)}{
\partial x^{j}}
\end{eqnarray}
which implies ${\mathcal F}^{ij}={\mathcal F}^{ij}(x)=F^{ij}(x)$, whereas the
second relation gives 
\begin{equation}
{{\mathcal M}}^{i}(x,\dot{x})={M}^{i}(x).
\end{equation}
The third provides 
\begin{equation}
\left[ \dot{x}^{i},{M}^{j}\right] =q\varepsilon
^{ij}{}_{k}A^{k}+q\varepsilon ^{j}{}_{kl}x^{k}\left[ \dot{x}
^{l},A^{i}\right] -q\varepsilon ^{jk}{}_{l}x_{k}F^{il},
\end{equation}
and the fourth becomes 
\begin{eqnarray}
\varepsilon _{ijk}{M}^{k} &=&\varepsilon _{ikl}x^{k}\left(
q\varepsilon _{jm}{}^{l}A^{m}-q\varepsilon _{jmn}x^{m}F^{ln}-q\left[
A^{l},L_{j}\right] \right)  \nonumber\\
&&-q\varepsilon _{jkl}x^{k}\left( \varepsilon _{im}{}^{l}A^{m}-q\varepsilon
_{imn}x^{m}F^{ln}-q\left[ A^{l},L_{i}\right] \right) \\
&&+q\varepsilon _{ikl}\varepsilon _{jmn}x^{k}x^{m}F^{ln}.\nonumber
\end{eqnarray}
The new ''Poincar\'{e} momentum'' solution of this last equation can be
expressed as 
\begin{equation}
{M}^{i}=q\varepsilon ^{i}{}_{jk}x^{j}A^{k}
\end{equation}
and then we recover the usual angular momentum 
\begin{equation}
\overrightarrow{{{\mathcal L}}}=\overrightarrow{L}+q\overrightarrow{r
}\wedge \overrightarrow{A}=\overrightarrow{r}\wedge \overrightarrow{p}.
\end{equation}

Note that in this approach we have no access to the Maxwell equations
because the Lie algebra of the linear momentum is trivial since $J\left(
p^{i},p^{j},p^{k}\right) =0$.

\subsection{Generalization to the curved space case}

The covariant Hamiltonian is now 
\begin{equation}
H=\frac{1}{2}m\,g_{ij}(x)\dot{x}^{i}\dot{x}^{j}.
\end{equation}
From the equation of motion we obtain as in the flat space case, the
following relation between bracket and metric tensor 
\begin{equation}
m\left[ x^{i},\dot{x}^{j}\right] =g^{ij}(x).
\end{equation}
We can generalize the notion of Poisson brackets between two functions
defined on a curved tangent bundled space by 
\begin{equation}
\left\{ f(x,\dot{x}),h(x,\dot{x})\right\} =
\frac{1}{m}g^{ij}(x) 
\left(
\frac{\partial f(x,\dot{x})}{\partial x^{i}}\frac{\partial h(x,
\dot{x})}{\partial \dot{x}^{j}}-\frac{\partial f(x,
\dot{x})}{\partial \dot{x}^{i}}\frac{\partial h(x,\dot{x})}{
\partial x^{j}}\right) .
\end{equation}
The others relations are obtained in a straightforward way 
\begin{equation}
\left\{ 
\begin{array}{l}
\D m\left[ x_{i},\dot{x}^{j}\right] =m\,g_{ik}\left\{ x^{k},\dot{x}
^{j}\right\} +m\,x^{k}\left\{ g_{ik},\dot{x}^{j}\right\} =\delta
_{i}{}^{j}+\frac{\partial g_{ik}}{\partial x_{j}}x^{k}, \\ 
\\ 
\D m\left[ x^{i},\dot{x}_{j}\right] =m\,g_{jk}\left\{ x^{i},\dot{x}
^{k}\right\} =\delta ^{i}{}_{j}, \\ 
\\ 
\D m\left[ x_{i},\dot{x}_{j}\right] =m\,g_{ik}g_{jl}\left\{ x^{k},
\dot{x}^{l}\right\} +m\,g_{jl}x^{k}\left\{ g_{ik},\dot{x}
^{l}\right\} =g_{ij}(x)+\frac{\partial g_{ik}}{\partial x^{j}}x^{k},
\end{array}
\right. 
\end{equation}
and the brackets between functions expandable in $x^{i}$ and $\dot{x}
^{i}$, have the form 
\begin{equation}
\left[ f(x,\dot{x}),h(x,\dot{x})\right] =\left\{
f(x,\dot{x}),h(x,\dot{x})\right\} +\frac{q}{m^{2}}{\mathcal F}
^{kl}\frac{\partial f(x,\dot{x})}{\partial \dot{x}^{k}}\frac{
\partial h(x,\dot{x})}{\partial \dot{x}^{l}},
\label{relation 1}
\end{equation}
where now the gauge curvature ${\mathcal F}^{kl}(x,\dot{x})$ is
velocity dependent 
\begin{equation}
\frac{q}{m^2}{\mathcal F}^{ij}(x,\dot{x})=\left[ g^{ik}\dot{x}
_{k},g^{jl}\dot{x}_{l}\right] =\frac{1}{m}\left( \frac{\partial
g^{ki}}{\partial x_{j}}-\frac{\partial g^{kj}}{\partial x_{i}}\right) 
\dot{x}_{k}+g^{ik}g^{jl}\left[ \dot{x}_{k},\dot{x}
_{l}\right] .  \label{relation 2}
\end{equation}

We define the angular momentum in a three dimensional curved space by the
usual relations 
\begin{equation}
\left\{ 
\begin{array}{l}
\D L_{i}=m\sqrt{g(x)}\varepsilon _{ijk}x^{j}\dot{x}^{k} \\ 
\phantom{L_{i}}=mE_{ijk}{}(x)x^{j}\dot{x}^{k}=mE_{i}{}^{jk}(x)x_{j}\dot{x}
_{k}, \\ 
\\ 
\D L^{i}=g^{ij}(x)L_{j}=m\sqrt{g(x)}g^{ij}(x)\varepsilon _{jkl}x^{k}
\dot{x}^{l} \\ 
\D\phantom{L^{i}}=\frac{m}{\sqrt{g(x)}}\varepsilon ^{ijk}{}x_{j}\dot{x}
_{k}=mE^{ijk}{}(x)x_{j}\dot{x}_{k},
\end{array}
\right.   \label{Ldef}
\end{equation}
where naturally $g(x)=\det \left( g_{ij}(x)\right) =\left( \det \left(
g^{ij}(x)\right) \right) ^{-1}$.

If we go back to the sO(3) symmetry laws, we obtain now 
\begin{equation}
\left\{ 
\begin{array}{l}
\D \left[ x^{i},L^{j}\right]=\left\{ x^{i},L^{j}\right\}
=E{}^{ij}{}_{k}x^{k}, \\ 
\\ 
\D \left[ \dot{x}^{i},L^{j}\right]=\left\{ \dot{x}^{i},L^{j}\right\}
+\frac{q}{m}E^{j}{}_{kl}x^{k}{\mathcal F}^{il} \\ 
\D\phantom{\left[ \dot{x}^{i},L^{j}\right]}=E^{ij}{}_{k}\dot{x}^{k}-\frac{1}{2}E^{j}{}_{kl}x^{k}\dot{x}
^{l}g_{mn}\frac{\partial g_{mn}}{\partial x_{i}}+\frac{q}{m}E^{j}{}_{kl}x^{k}
{\mathcal F}^{il}, \\ 
\\ 
\D \left[ L^{i},L^{j}\right]=\left\{ L^{i},L^{j}\right\}
+qE{}^{i}{}_{kl}E{}^{j}{}_{mn}x^{k}x^{m}{\mathcal F}^{ln} \\ 
\D\phantom{\left[ L^{i},L^{j}\right]}=E^{ij}{}_{k}L^{k}+
\frac{1}{2}E^{i}{}_{kl}E^{j}{}_{mn}x^{k}x^{m}\left( 
\dot{x}^{l}g_{pq}\frac{\partial g^{pq}}{\partial x_{n}}-
\dot{x}^{n}g_{pq}\frac{\partial g^{pq}}{\partial x_{l}}\right)  \\ 
\D\phantom{\left[ L^{i},L^{j}\right]=} 
+qE_{kl}^{i}E{}^{j}{}_{mn}x^{k}x^{m}{\mathcal F}^{ln} \\ 
\D\phantom{\left[ L^{i},L^{j}\right]=} 
+E_{qmn}x^{k}x^{m}\dot{x}^{n}\left( E^{j}{}_{kl}\frac{\partial g^{iq}
}{\partial x_{l}}-E^{i}{}_{kl}\frac{\partial g^{jq}}{\partial x_{l}}\right) ,
\end{array}
\right. 
\end{equation}
where as usual we have used the formula $\frac{\partial g(x)}{\partial x^{i}
}=g(x)g_{jk}(x)\frac{\partial g^{jk}(x)}{\partial x^{i}}$ .

We shall use two examples of curvatures. The first is the standard gauge
electromagnetic curvature, the second comes from the electromagnetic type
gravity in the parametrized post newtonian formalism (the so-called ''PPN
formalism'') that we will see in a future section.

\subsubsection{Without connection}

In order to restore the sO(3) symmetry as in the ''flat case'' we choose
the transformation law of the angular momentum 
\begin{equation}
L^{i}\rightarrow {\mathcal L}^{i}=L^{i}+{\mathcal M}^{i}(x,\dot{x}),
\end{equation}
and we impose as usual the relations 
\begin{equation}
\left\{ 
\begin{array}{c}
\left[ x^{i},{\mathcal L}^{j}\right]=E{}{}^{ij}{}_{k}x^{k}, \\ 
\\ 
\left[ \dot{x}^{i},{\mathcal L}^{j}\right]=E{}^{ij}{}_{k}\dot{x}
^{k}, \\ 
\\ 
\left[ {\mathcal L}^{i},{\mathcal L}^{j}\right]=E{}^{ij}{}_{k}{\mathcal L}^{k}.
\end{array}
\right.   \label{relwc}
\end{equation}
The first equation in (\ref{relwc}) implies as in the flat space case that
the new angular momentum is velocity independent, so we note ${\mathcal M}
^{i}(x,\dot{x})=M^{i}(x)$. The second equation in (\ref{relwc})
gives 
\begin{eqnarray}
\left[ \dot{x}^{i},M^{j}\right]  &=&-\frac{1}{2mg}\frac{\partial g}{
\partial x_{i}}L^{j}-\frac{q}{m}E_{\;lm}^{j}x^{l}{\mathcal F}^{im}(x,
\dot{x})  \label{crochetxpM} \\
&=&-\frac{1}{2mg}\frac{\partial g}{\partial x_{i}}L^{j}-\frac{1}{m}\left( 
\frac{\partial g^{ki}}{\partial x_{m}}-\frac{\partial g^{km}}{\partial x_{i}}
\right) \dot{x}_{k}E_{\;lm}^{j}x^{l}  \nonumber \\
&\phantom{=}&-\frac{q}{m}E_{\;lm}^{j}x^{l}g^{ik}g^{ml}\left[ \dot{x}
_{k},\dot{x}_{l}\right].\nonumber
\end{eqnarray}
Assuming the Jacobi identity 
\begin{equation}
J\left( x^{i},\dot{x}^{j},\dot{x}^{k}\right) =\frac{q}{m^{3}}
\frac{\partial {\mathcal F}^{jk}(x,\dot{x})}{\partial \dot{x}
_{i}}+\frac{1}{m^2}\left( \frac{\partial g^{ki}}{\partial x_{j}}-\frac{
\partial g^{ji}}{\partial x_{k}}\right) =0,  \label{Jacobint}
\end{equation}
implies that the term 
$\frac{q}{m^2}F^{ij}\left( x\right) =g^{ik}g^{il}\left[ \dot{x}_{k},
\dot{x}_{l}\right]$ in (\ref{relation 2}) is velocity independent whereas 
the whole gauge field ${\mathcal F}^{ij}$ is velocity dependent. 
Now, since $M^{i}$ is velocity
independent, the velocity dependent part of the left hand side of the
equation (\ref{crochetxpM}) must vanish leading to the relation 
\begin{equation}
\frac{\partial g}{\partial x_{i}}L^{j}=-2g\displaystyle \left( \displaystyle
\frac{\partial g^{ki}}{\partial x_{m}}-\displaystyle\frac{\partial g^{km}}{
\partial x_{i}}\right) \dot{x}_{k}E_{\;lm}^{j}x^{l}.  \label{Limodif}
\end{equation}
If we compare (\ref{Limodif}) with the definition of $L^{j}$ (\ref{Ldef}) we
obtain a constraint relation on the metric tensor 
\begin{equation}
\frac{\partial g}{\partial x_{i}}g^{nk}=-2g\displaystyle \left( \displaystyle
\frac{\partial g^{ki}}{\partial x_{n}}-\displaystyle\frac{\partial g^{kn}}{
\partial x_{i}}\right) \,,\forall i\neq n,  \label{relation 3}
\end{equation}
that we will not use in the following.

Equation (\ref{crochetxpM}) reads now 
\begin{equation}
\left[ \dot{x}^{i},M^{j}\right] =-\frac{q}{m}E_{\;lm}^{j}x^{l}F^{im},
\end{equation}
leading to the same result as in the flat space case, that is, 
the angular momentum $M$ is a Poincar\'{e} momentum equal to 
\begin{equation}\label{MM}
M^{i}(x)=\frac{1}{2}qE_{jkl}x^{i}x^{k}F^{jl}=-q\left( \overrightarrow{r}
\overrightarrow{B}\right) x^{i}.
\end{equation}
This relation (\ref{MM}) 
still implies the presence of a Dirac magnetic monopole 
\begin{equation}
\overrightarrow{B}=\frac{g}{4\pi }\frac{\overrightarrow{r}}{r^{3}}.
\end{equation}
The result of this computation is that the sO(3) symmetry algebra, in flat
as well as in curved space, is restored by introducing the same Dirac
monopole.

In (\ref{crochetxpM}) we have used the relation 
\begin{equation}
\frac{q}{m^2}{\mathcal F}^{ij}(x,\dot{x})=\frac{1}{m}\left( \frac{
\partial g^{ki}}{\partial x_{j}}-\frac{\partial g^{kj}}{\partial x_{i}}
\right) \dot{x}_{k}+\frac{q}{m^2}F^{ij}(x),  \label{relFF}
\end{equation}
which shows the link between the non abelian gauge curvature ${\mathcal F}
^{ij}(x,\dot{x})$ and the usual electromagnetic type abelian gauge
curvature $F^{ij}(x)$. Now, we are also able to recover the equation of
motion of a particle in a curved space and in the presence of an
electromagnetic field. Indeed from (\ref{metrique}) and (\ref{relFF}) we get
the well known equation of motion 
\begin{equation}
m\ddot{x}^{i}=-m\Gamma _{jk}^{i}(x)\dot{x}^{j}\dot{x}
^{k}+qF^{ij}(x)\dot{x}_{j},
\end{equation}
where we have introduced the standard Christoffel symbols $\Gamma _{jk}^{i}$.

\subsubsection{With connection}

We use the two transformation laws and we choose the ''Poincar\'{e}
momentum'' as in the abelian case 
\begin{equation}
\left\{ 
\begin{array}{l}
\dot{x}^{i}\rightarrow p^{i}=m\dot{x}^{i}+q{\mathcal A}^{i}(x,
\dot{x}), \\ 
\\ 
L^{i}\rightarrow {\mathcal L}^{i}=L^{i}+{\mathcal M}^{i}(x,\dot{x}) \\ 
\phantom{L^{i}\rightarrow {\mathcal L}^{i}}
=L^{i}+\left( \overrightarrow{r}\wedge q\overrightarrow{{\mathcal A}}{\mathcal 
(}x,\dot{x})\right) ^{i}=\left( \overrightarrow{r}\wedge 
\overrightarrow{p}\right) ^{i},
\end{array}
\right.   \label{linear momentum}
\end{equation}
then 
\begin{equation}
\left[ x^{i},p^{j}\right]=g^{ij}(x)+\frac{q}{m}\frac{\partial 
{\mathcal A}^{j}(x,
\dot{x})}{\partial \dot{x}_{i}}=G^{ij}(x,\dot{x}).
\end{equation}
We note that in this context the spatial ''metric tensor'' $G^{ij}(x,
\dot{x})$ is also velocity dependent. If we require the relations 
\begin{equation}
\left\{ 
\begin{array}{l}
\left[ p^{i},p^{j}\right]=\left\{ p^{i},p^{j}\right\} =0, \\ 
\\ 
\left[ x^{i},{\mathcal L}^{j}\right]=\left\{ x^{i},{\mathcal L}^{j}\right\}
=E^{ij}{}_{k}x^{k}, \\ 
\\ 
\left[ p^{i},{\mathcal L}^{j}\right]=\left\{ p^{i},{\mathcal L}^{j}\right\}
=E^{ij}{}_{k}p^{k}, \\ 
\\ 
\left[ {\mathcal L}^{i},{\mathcal L}^{j}\right]=\left\{ {\mathcal L}^{i},{\mathcal L}
^{j}\right\} =E^{ij}{}_{k}{\mathcal L}^{k},
\end{array}
\right. 
\end{equation}
the first relation gives 
\begin{eqnarray}
{\mathcal F}^{ij}(x,\dot{x}) &=&m\left[ \dot{x}^{j},{\mathcal A
}^{i}(x,\dot{x})\right] -m\left[ \dot{x}^{i},{\mathcal A}
^{j}(x,\dot{x})\right] +\left[ {\mathcal A}^{i}(x,\dot{x}),
{\mathcal A}^{j}(x,\dot{x})\right]   \nonumber \\
&=&\frac{\partial {\mathcal A}^{j}(x,\dot{x})}{\partial x^{i}}-\frac{\partial
{\mathcal A}^{i}(x,\dot{x})}{\partial x^{j}}  \nonumber \\
&&+{\mathcal F}^{jk}(x,\dot{x})\frac{\partial {\mathcal A}^{i}(x,\dot{x})
}{\partial \dot{x}^{k}}-{\mathcal F}^{ik}(x,\dot{x})\frac{
\partial {\mathcal A}^{j}(x,\dot{x})}{\partial \dot{x}^{k}}  \nonumber
\\
&&+\left[ {\mathcal A}^{i}(x,\dot{x}),{\mathcal A}^{j}(x,
\dot{x})\right] .
\end{eqnarray}
The third equation implies 
\begin{equation}
\frac{1}{2}g_{kl}\frac{\partial g^{kl}}{\partial x_{i}}{\mathcal L}
^{j}+E^{j}{}_{kl}{}p^{l}\left[ {\mathcal A}^{i}{\mathcal (}x,\dot{x}
),x^{k}\right] +E^{j}{}_{kl}x^{k}p^{l}\left[ {\mathcal A}^{i}(x,
\dot{x}),\sqrt{g}\right] =0,
\end{equation}
and the fourth one gives
\begin{eqnarray}
&&\frac{1}{2}x^{m}g_{kl}\frac{\partial g^{kl}}{\partial x_{n}}\left(
E^{i}{}_{mn}{\mathcal L}^{j}-E^{j}{}_{mn}{\mathcal L}^{i}\right)   \nonumber \\
&&+\varepsilon _{kl}^{i}E^{j}{}_{mn}x^{k}x^{m}\left( \left[ \sqrt{g},
{\mathcal A}^{n}{\mathcal (}x,\dot{x})\right] p^{l}-\left[ \sqrt{g},
{\mathcal A}^{l}{\mathcal (}x,\dot{x})\right] p^{n}\right)  \nonumber\\
&&+E^{i}{}_{kl}E^{j}{}_{mn}\left( \left[ x^{k},{\mathcal A}^{n}{\mathcal (}x,
\dot{x})\right] x^{m}p^{l}-\left[ x^{m},{\mathcal A}^{l}{\mathcal (}x,
\dot{x})\right] x^{k}p^{n}\right) =0.
\end{eqnarray}

We can easily check that the last two equations are compatible with the
chosen form for the generalized angular momentum (\ref{linear momentum}).
These last two relations must be seen as definitions of the non abelian
connexion from the metric tensor.

\section{Lorentz algebra}

The natural extension of the angular algebra is obviously the Lorentz
algebra which is primordial in the goal to envisage the gravitation theory
in the frame of general relativity. Here we study only the case without the
connection field.

\subsection{Non abelian case}

In the curved quadri-dimensional space we obtain for the fundamental
bra\-cket relations 
\begin{equation}
\left\{ 
\begin{array}{l}
\D m\left[ x^{\mu },\dot{x}^{\nu }\right] =g^{\mu \nu }(x), \\ 
\\ 
\D m\left[ x_{\mu },\dot{x}_{\nu }\right] =g_{\mu \nu }(x)+x^{\rho }
\frac{\partial g_{\mu \rho }(x)}{\partial x^{\nu }}, \\ 
\\ 
\D m\left[ \dot{x}^{\mu },\dot{x}^{\nu }\right] =\frac{q}{m}
{\mathcal F}^{\mu \nu }(x,\dot{x}), \\ 
\\ 
\D m\left[ \dot{x}_{\mu },\dot{x}_{\nu }\right] =\frac{q}{m}
F_{\mu \nu }(x).
\end{array}
\right. 
\end{equation}
It is convenient to define the angular quadri-momentum under the form 
\begin{equation}
L_{\mu \nu }=m\left( x_{\mu }\dot{x_{\nu }}-x_{\nu }\dot{x}
_{\mu }\right) ,
\end{equation}
which gives the deformed Lorentz algebra with the following structure law 
\begin{equation}
\left\{ 
\begin{array}{l}
 \left[ x_{\mu },L_{\rho \sigma }\right] =\left\{ x_{\mu },L_{\rho \sigma
}\right\} =g_{\mu \sigma }x_{\rho }-g_{\mu \rho }x_{\sigma }
+x_{\rho}x^{\lambda }\frac{\partial g_{\mu \lambda }(x)}{\partial x^{\sigma }}
-x_{\sigma }x^{\lambda }\frac{\partial g_{\mu \lambda }(x)}{\partial x^{\rho
}}, \\ 
\\ 
 \left[ \dot{x}_{\mu },L_{\rho \sigma }\right] =\left\{ \dot{
x_{\mu }},L_{\rho \sigma }\right\} +\frac{q}{m}(F_{\mu \sigma }\dot{
x_{\rho }}-F_{\mu \rho }\dot{x_{\sigma }}) \\
=g_{\mu \sigma }\dot{x_{\rho }}-g_{\mu \rho }\dot{x_{\sigma }
}+\dot{x}_{\rho }x^{\lambda }\frac{\partial g_{\mu \lambda }(x)}{
\partial x^{\sigma }}-\dot{x}_{\sigma }x^{\lambda }\frac{\partial
g_{\mu \lambda }(x)}{\partial x^{\rho }}+\frac{q}{m}(F_{\mu \sigma }
\dot{x_{\rho }}-F_{\mu \rho }\dot{x_{\sigma }}), \\ 
\\ 
 \left[ L_{\mu \nu },L_{\rho \sigma }\right] =\left\{ L_{\mu \nu },L_{\rho
\sigma }\right\} +q(x_{\mu }x_{\rho }F_{\nu \sigma }-x_{\nu }x_{\rho }F_{\mu
\sigma }+x_{\mu }x_{\sigma }F_{\rho \nu }-x_{\nu }x_{\sigma }F_{\rho \mu })
\\ 
 =g_{\mu \rho }L_{\nu \sigma }-g_{\nu \rho }L_{\mu \sigma }+g_{\mu \sigma
}L_{\rho \nu }-g_{\nu \sigma }L_{\rho \mu } \\ 
 +m\left( x_{\rho }\dot{x}_{\nu }x^{\lambda }\frac{\partial
g_{\lambda \sigma }}{\partial x^{\mu }}-x_{\nu }\dot{x}_{\rho
}x^{\lambda }\frac{\partial g_{\lambda \mu }}{\partial x^{\sigma }}+x_{\mu }
\dot{x}_{\rho }x^{\lambda }\frac{\partial g_{\lambda \mu }}{\partial
x^{\sigma }}-x_{\rho }\dot{x}_{\mu }x^{\lambda }\frac{\partial
g_{\lambda \sigma }}{\partial x^{\mu }}\right.  \\ 
 \left. +x_{\nu }\dot{x}_{\sigma }x^{\lambda }\frac{\partial
g_{\lambda \rho }}{\partial x^{\mu }}-x_{\sigma }\dot{x}_{\nu
}x^{\lambda }\frac{\partial g_{\lambda \mu }}{\partial x^{\rho }}+x_{\sigma }
\dot{x}_{\mu }x^{\lambda }\frac{\partial g_{\lambda \nu }}{\partial
x^{\rho }}-x_{\mu }\dot{x}_{\sigma }x^{\lambda }\frac{\partial
g_{\lambda \rho }}{\partial x^{\nu }}\right)  \\ 
 +q(x_{\mu }x_{\rho }F_{\nu \sigma }-x_{\nu }x_{\rho }F_{\mu \sigma }+x_{\mu
}x_{\sigma }F_{\rho \nu }-x_{\nu }x_{\sigma }F_{\rho \mu }).
\end{array}
\right. 
\end{equation}

In the same spirit, as in the sO(3) algebra, we choose to restore the
Lorentz symmetry with the following angular quadri-momentum transformation
law 
\begin{equation}
L_{\mu \nu }\rightarrow {\mathcal L}_{\mu \nu }=L_{\mu \nu }+{\mathcal M}_{\mu
\nu }(x,\dot{x}),
\end{equation}
and we require the usual structure 
\begin{equation}
\left\{ 
\begin{array}{lll}
\left[ x_{\mu },{\mathcal L}_{\rho \sigma }\right]  &=&\left\{ x_{\mu },
{\mathcal L}_{\rho \sigma }\right\} =g_{\mu \sigma }x_{\rho }-g_{\mu \rho
}x_{\sigma },  \label{lie1} \\
&&  \nonumber \\
\left[ \dot{x}_{\mu },{\mathcal L}_{\rho \sigma }\right]  &=&\left\{ 
\dot{x}_{\mu },{\mathcal L}_{\rho \sigma }\right\} =g_{\mu \sigma }
\dot{x}_{\rho }-g_{\mu \rho }\dot{x}_{\sigma },  \label{lie3}
\\
&&  \nonumber \\
\left[ {\mathcal L}_{\mu \nu },{\mathcal L}_{\rho \sigma }\right]  &=&\left\{ 
{\mathcal L}_{\mu \nu },{\mathcal L}_{\rho \sigma }\right\} =g_{\mu \rho }
{\mathcal L}_{\upsilon \sigma }-g_{\nu \rho }{\mathcal L}_{\mu \sigma }+g_{\mu
\sigma }{\mathcal L}_{\rho \nu }-g_{\nu \sigma }{\mathcal L}_{\rho \mu }.
\end{array}
\right.
\label{lie5}
\end{equation}
From (\ref{lie1}) we easily deduce that the quadri-momentum
${\mathcal M}_{\mu \nu }(x,\dot{x})$ is 
only position dependent, 
${\mathcal M}_{\mu \nu }(x,\dot{x})=M_{\mu \nu }(x)$. We call this
quadri-tensor the Poincar\'{e} tensor. Equation (\ref{lie3}) then gives 
\begin{equation}
\left[ \dot{x}_{\mu },M_{\rho \sigma }\right] =\frac{q}{m}(F_{\mu
\sigma }x_{\rho }-F_{\mu \rho }x_{\sigma }).  \label{neufter}
\end{equation}
This result (\ref{neufter}) together with the third relation given in
(\ref{lie5}) implies
\begin{eqnarray}
&&g_{\mu \rho }M_{\nu \sigma }-g_{\nu \rho }M_{\mu \sigma }+g_{\mu \sigma
}M_{\rho \nu }-g_{\nu \sigma }M_{\rho \mu }  \nonumber \\
&=&q(F_{\nu \sigma }x_{\mu }x_{\rho }-F_{\mu \sigma }x_{\nu }x_{\rho
}+F_{\rho \nu }x_{\mu }x_{\sigma }-F_{\rho \mu }x_{\nu }x_{\sigma }).
\label{neuf}
\end{eqnarray}

In order to determine the Poincar\'{e} tensor, let us firstly consider the
case $\nu =\sigma =i$, where $i=1,2,3$, and sum over $i$. Equation (\ref
{neuf}) becomes 
\begin{equation}
-g^{i}{}_{\rho }M_{\mu i}+g_{\mu }{}^{i}M_{\rho i}-3M_{\rho \mu }=q(-F_{\mu
i}x^{i}x_{\rho }+F_{\rho i}x_{\mu }x^{i}-F_{\rho \mu }r^{2}),
\label{neufbis}
\end{equation}
and with $\rho =j$ and $\mu =k$, we obtain 
\begin{equation}
M_{ij}=q(F_{ij}x^{k}x_{k}-F{}_{jk}x^{k}x_{i}-F_{ki}{}x^{k}x_{j}).
\label{Equation Moment}
\end{equation}
Using the definition $M_{i}=\varepsilon {}_{i}{}^{jk}M_{jk}$ , the same
magnetic angular momentum as for the $sO(3)$ case is deduced as expected for
the spatial degrees of freedom ($i=1,2,3$) 
\begin{equation}
\overrightarrow{M}=-q(\overrightarrow{r}.\overrightarrow{B})\overrightarrow{r
}.  \label{Champ Magnetique}
\end{equation}
Equation (\ref{neufter}) becomes then 
\begin{equation}
\left\{ 
\begin{array}{l}
\left[ \dot{x}_{k},M_{ij}\right] =\frac{q}{m}\left(
F_{kj}x_{i}-F_{ki}x_{j}\right) , \\ 
\\ 
\left[ \dot{x}_{0},M_{ij}\right] =\frac{q}{m}\left(
F_{0j}x_{i}-F_{0i}x_{j}\right) ,
\end{array}
\right.   \label{eq1}
\end{equation}
and if we introduce the magnetic angular momentum (\ref{Champ Magnetique})
into the set of equations (\ref{eq1}) we obtain 
\begin{equation}
\left\{ 
\begin{array}{l}
x_{i}B_{j}+x_{j}B_{i}=-x_{j}x^{k}\frac{\partial B_{k}}{\partial x^{i}}, \\ 
\\ 
F_{0j}x_{i}-F_{0i}x_{j}=\left( \overrightarrow{r}\wedge \overrightarrow{E}
\right) _{k}=0,
\end{array}
\right.   \label{relation}
\end{equation}
which has radial vector fields centered at the origin as solutions 
\begin{equation}
\left\{ 
\begin{array}{l}
\D\overrightarrow{B}=\frac{g}{4\pi }\frac{\overrightarrow{r}}{
r^{3}}, \\ 
\\ 
\overrightarrow{E}=q^{\prime }f(r)\overrightarrow{r}.
\end{array}
\right. 
\end{equation}
Then we are in presence of a Schwinger dyon of magnetic charge $g$ and
electric charge $q^{\prime }$, a priori different from $q$ the electric
charge of the particle.

We still obtain the standard Poincar\'{e} magnetic angular momentum 
\cite{POINCARE} 
\begin{equation}
\overrightarrow{M}=-\frac{qg}{4\pi }\frac{\overrightarrow{r}
}{r},
\end{equation}
and naturally we can point out that this Poincar\'{e} angular momentum has a
constant norm and as Dirac in the quantum theory context we can put this
norm equal to $nh$.

Secondary we consider the ''boost'' part of (\ref{neuf}) - for $\rho =0$,
and $\mu =j$, in the equation (\ref{neufbis}) we have for the temporal
components of the Poincar\'{e} tensor 
\begin{equation}
{\mathcal M}_{0j}=q(-F_{ji}x^{i}x_{0}-F_{0i}x_{j}x^{i}+F_{0j}r^{2}),
\end{equation}
using now the Poincar\'{e} magnetic momentum we find for its component 
\begin{equation}
{\mathcal M}_{0j}{\mathcal =}q\left[ -\left( \overrightarrow{r}\wedge 
\overrightarrow{B}\right) _{j}x_{0}-\left( \overrightarrow{r}\overrightarrow{
E}\right) x_{j}+r^{2}E_{j}\right] =0.
\end{equation}

If we require the following Jacobi identity 
\begin{equation}
J\left( \dot{x}_{\mu },\dot{x}_{\nu },\dot{x}_{\rho
}\right) =\frac{q}{m^3}
\left(\frac{\partial F_{\mu \nu }(x)}{\partial x^{\rho }}+\frac{\partial
F_{\nu \rho }(x)}{\partial x^{\mu }}+\frac{\partial F_{\rho \mu }(x)}{
\partial x^{\nu }}\right)=0,
\end{equation}
we retrieve in particular that $div\overrightarrow{B}=0$,
then the Poincar\'{e} momentum is zero.

\underline{Remark:} If we introduce the linear momentum $p^{\mu }$ , the
computation is the same as in the previous part with 
\begin{equation}
\left\{ 
\begin{array}{c}
\dot{x}^{\mu }\rightarrow p^{\mu }=m\dot{x}^{\mu }+q{\mathcal A}
^{\mu }(x,\dot{x}), \\ 
\\ 
L^{\mu \nu }\rightarrow {\mathcal L}^{\mu \nu }=L^{\mu \nu }+{\mathcal M}^{\mu
\nu }(x) \\ 
=x^{\mu }p^{\nu }-x^{\nu }p^{\mu }.
\end{array}
\right. 
\end{equation}

\subsection{Brief application to gravitoelectromagnetism}

It is well known that for interpreting the experimental tests of gravitation
theories, the Parametrized Post-Newtonian formalism (PPN) is often used 
\cite{WHEELER}, where the limit of low velocities and small stresses is 
taken. In
this formalism, gravity is described by a general type metric which contains
dimensionless constants call PPN-parameters, that are powerful tools in
theoretical astrophysics. This formalism was applied by Braginski {\it et al}
\cite{BRAGINSKI} to propose laboratory experiments to test relativistic gravity
and in particular to study gravitoelectromagnetism. They analyzed
magnetic and electric type gravity using a truncated and rewritten version
of the PPN formalism by deleting certain parameters not present in general
relativity and all gravitational non linearities. Recently Mashhoon 
wrote a theoretical paper \cite{MASHHOON} where he considers several
important quantities relative to this theory like field equations,
gravitational Larmor theorem or stress-energy tensor. He introduced
gravitoelectromagnetism which is based upon the formal analogy between
gravitational Newton potential and electric Coulomb potential. A long time
ago Holzmuller \cite{HOLZMULLER} and Tisserand \cite{TISSERAND} have already
postulated gravitational electromagnetic components for the gravitational
influence of the sun on the motion of planets. Finally Mashhoon \cite{MASHHOON}
considers that a particle of inertial mass m has gravitoelectric charge $
q_{E}=-m$ and gravitomagnetic ch$\arg $e $q_{M}=-2m$ , the numerical factor
2 coming from the spin character of the gravitational field. In the final
part of this work we apply these last ideas to our formalism.

Suppose that gravitation creates a gravitoelectromagnetic field, then we
have the following equation

\begin{equation}
{\mathcal M}
_{ij}=q(F_{ij}x^{k}x_{k}-F_{jk}x^{k}x_{i}-F_{ki}x^{k}x_{j})+
g(^*\!F_{ij}x^kx_k-^*\!F_{jk}x^kx_{i}-^*\!F_{ki}x^kx_j).
\end{equation}
Here we introduce the Hodge duality such as
\begin{equation}
\left[ \dot{x}_{\mu },\dot{x}_{\nu }\right] =-\frac{1}{m^{2}}
(qF_{\mu \nu }+g^{*}\!F_{\mu \nu }),
\end{equation}
$g$ being the magnetic charge of the particle, which can be seen now as a
Schwinger dyon.

The new angular momentum is the sum of two contributions, a gravitomagnetic
and a gravitoelectric one
\begin{equation}
\overrightarrow{M}=-q(\overrightarrow{r}.\overrightarrow{B})\overrightarrow{r
}+g(\overrightarrow{r}.\overrightarrow{E})\overrightarrow{r}=\overrightarrow{
M_{m}}+\overrightarrow{M_{e}},  \label{vingt deux}
\end{equation}
where
\begin{equation}
\left\{ 
\begin{array}{c}
\overrightarrow{M_{m}}=-q(\overrightarrow{r}.\overrightarrow{B})
\overrightarrow{r}, \\ 
\\ 
\overrightarrow{M_{e}}=g(\overrightarrow{r}.\overrightarrow{E})
\overrightarrow{r}
\end{array}
\right. 
\end{equation}
are the gravitomagnetic and gravitoelectric angular momenta.
Due to the fact that for these gravitational monopoles the source of the
fields is localized at the origin, we obtain for the vector 
$\overrightarrow{P}=q\overrightarrow{B}-g\overrightarrow{E}$
\begin{eqnarray}
div\overrightarrow{P} &=&m^3J\left( \dot{x}^{i},\dot{x}^{j},\nonumber
\dot{x}^{k}\right) =q\,div\overrightarrow{B}-g\,div\overrightarrow{E} \\
&=&\frac{qg}{4\pi }\left[x^{l},\frac{x_{l}}{r^{3}}\right]=qg\,\delta ^{3}(
\overrightarrow{r}).
\end{eqnarray}
So as an example 
\begin{equation}
\left\{ 
\begin{array}{c}
\D\overrightarrow{B}=\frac{g^{\prime }}{4\pi }\frac{\overrightarrow{r}}{r^{3}},
\\ 
\\ 
\D\overrightarrow{E}=-\frac{q^{\prime }}{4\pi }\frac{\overrightarrow{r}}{r^{3}}
,
\end{array}
\right. 
\end{equation}
where a priori  $g^{\prime }$ and $q^{\prime }$ are different from $g$ and $
q$.
If we require now the Jacobi identity between the velocities
\begin{equation}
J\left( \dot{x}^{\mu },\dot{x}^{\nu },\dot{x}^{\rho }
\right) =0,  \label{quinze}
\end{equation}
we have the generalized gravitational Maxwell equations 
\begin{equation}
q(\partial ^{\mu }F^{\nu \rho }+\partial ^{\nu }F^{\rho \mu }+\partial
^{\rho }F^{\mu \nu })+g(\partial ^{\mu }{}^{*}\!F^{\nu \rho }+\partial ^{\nu
}{}^{*}\!F^{\rho \mu }+\partial ^{\rho }{}^{*}\!F^{\mu \nu })=0.
\label{quatorze}
\end{equation}
The projection of (\ref{quatorze}) on three dimensional space gives
\begin{equation}
q\,div\overrightarrow{B}-g\,div\overrightarrow{E}=div \overrightarrow{P}=0,
\end{equation}
where $\overrightarrow{P}$ can be taken either perpendicular to the
vector $\overrightarrow{r}$ or null, we then have $\overrightarrow{M}=
\overrightarrow{0}.$

If we don't require this identity, we have
\begin{equation}
q(\partial ^{\mu }F^{\nu \rho }+\partial ^{\nu }F^{\rho \mu }+\partial
^{\rho }F^{\mu \nu })+g(\partial ^{\mu }{}^{*}\!F^{\nu \rho }+\partial ^{\nu
}{}^{*}\!F^{\rho \mu }+\partial ^{\rho }{}^{*}\!F^{\mu \nu })=qgN^{\mu \nu \rho
},  \label{dixsept}
\end{equation}
where $N^{\mu \nu \rho }$ is the components of a 3-differential form $N$
which breaks this identity.

In both cases (Jacobi identity required or not) we obtain the following two
groups of gravitational Maxwell equations \cite{NOUS1}
\begin{equation}
\left\{ 
\begin{array}{c}
\delta F=j, \\ 
\\ 
dF=-^{*}k,
\end{array}
\right. 
\end{equation}
but in the case where the Jacobi identity is not required we have the
relation $g^{*}j-q^{*}k=qgN$ , and for the case where the Jacobi identity
is required this relation is replaced by  $g^{*}j=q^{*}k.$

Then in the frame of gravitoelectromagnetism we have for the two charges 
($q,g$) of the
dyon particle and for the two charges ($q^{\prime },g^{\prime }$) of the
gravitoelectromagnetic dyon similar to the one studied by Mashhoon 
\cite{MASHHOON}
\begin{equation}
\left\{ 
\begin{array}{ccc}
\D q&=&\alpha \D\frac{q_{E}}{\varepsilon _{0}}, \\ 
\\ 
\D g&=&\beta \mu _{0}q_{M}, \\ 
\\ 
\D q^{\prime }&=&\alpha ^{\prime }\D\frac{q_{E}^{\prime }}{\varepsilon _{0}},
\\ 
\\ 
\D g^{\prime }&=&\beta ^{\prime }\mu _{0}q_{M}^{\prime },
\end{array}
\right. 
\end{equation}
where $\varepsilon _{0}$ and $\mu _{0}$ are respectively the vacuum
permittivity and permeability, and $\alpha ,\beta ,\alpha ^{\prime }$ and $
\beta ^{\prime }$ are constants with the following dimensional relations
\begin{equation}
\left\{ 
\begin{array}{ccccc}
\left[ \alpha \right] &=&\left[ \alpha ^{\prime }\right] &=&L^{-2}, \\ 
\\ 
\left[ \beta \right] &=&\left[ \beta ^{\prime }\right] &=&L^{-1}T^{-1}.
\end{array}
\right. 
\end{equation}
From the quantization of the relation (\ref{vingt deux}) we deduce
\begin{equation}
\alpha \beta ^{\prime }q_{E}q_{M}^{\prime }+\alpha ^{\prime }\beta
q_{E}^{\prime }q_{M}=\frac{2nh}{Z^2_{0}},
\end{equation}
where $Z_{0}$ is the vacuum impedance, and if we postulate as in
Mashhoon's paper \cite{MASHHOON} the following relations 
\begin{equation}
\left\{ 
\begin{array}{ccc}
q_{E}&=&a\sqrt{G}m, \\ 
\\ 
q_{M}&=&b\sqrt{G}m, \\ 
\\ 
q_{E}^{\prime }&=&a^{\prime }\sqrt{G}m^{\prime }, \\ 
\\ 
q_{M}^{\prime }&=&b^{\prime }\sqrt{G}m^{\prime },
\end{array}
\right. 
\end{equation}
where $a,b,a^{\prime}$ and $b^{\prime}$ are constants, 
$m^{\prime }$ is the monopole mass {\it a priori}
 different from the mass $m$ of the
particle and $G$ is the gravitational constant. We have also $\frac{q_{M}}{
q_{E}}=\frac{q_{M}^{\prime }}{q_{E}^{\prime }}=\frac{b}{a}=\frac{b^{\prime }
}{a^{\prime }}=s$ which is the Mashhoon's 
relation between electric and magnetic
charges and the spin of the gauge boson interaction. In the
gravitoelectromagnetic theory we naturally have $s=2$ , then we deduce
\begin{equation}
m\,m^{\prime }=\frac{nh}{aa^{\prime }(\alpha \beta ^{\prime }+\alpha
^{\prime }\beta )GZ^2_{0}}=\frac{nh}{AGZ^2_{0}},
\end{equation}
where $A$ is a new constant and $n$ is an integer number in the Schwinger
formalism (bosonic spectrum) and half integer number in the Dirac formalism
(fermionic and bosonic spectrum). We have then obtain a qualitative relation
which gives the mass spectrum of dyon particles and, in the particular case
where $q=q^{\prime }$ and $g=g^{\prime }$ , it simply becomes
\begin{equation}
m=m^{\prime }=\sqrt{\frac{nh}{AGZ^2_{0}}}.
\end{equation}

\section{Conclusion}

We have studied the breaking of sO(3) and Lorentz algebras symmetries
by ''abelian'' or
''non abelian'' gauge curvature in a covariant formalism. The restoration of
these algebra symmetries is made in two different ways: the first
one with the introduction of a generalized angular momentum which is the
Poincar\'{e} momentum and the second one with a Legendre transformation of
the velocity that naturally introduces a linear  momentum and a connexion.
We pointed out that it is more interesting to work in the tangent bundle
approach if we want to study the equations of motion and the existence of a
monopole than in the cotangent bundle approach. The application of this
formalism to the gravitoelectromagnetism theory was envisaged. The principal
result of this last part is the extrapolation of the Dirac quantum condition
for the magnetic monopole to the case of a potential gravitoelectromagnetic
monopole. Finally using Mashhoon's postulate for the relation between the
gravitocharges and the mass of the particle, we obtain a qualitative
condition on the mass spectrum.

Actually we look at the theory of spinning point particles in different
terms than Chou \cite{CHOU} , then we retrieve in particular the principal
results that Van Holten \cite{VANHOLTEN} has found in an other context.

\section*{Acknowledgments} 

A.B. would like to thank Patrice P\'{e}rez for helpful discussions.

\end{document}